%% file: paper.tex
\documentclass{elsart}
\usepackage{amsmath,graphicx}
\usepackage{amsmath,feynmp}
\usepackage{epsfig}
\usepackage{rotate}
\usepackage{lscape}
\usepackage{feynarts}
\usepackage{bm}

\allowdisplaybreaks

\def\P{\mathcal P}

\def\dd{{\mathrm d}}
\def\ln{{\mathrm ln}}

\def\fun#1#2{\lower3.6pt\vbox{\baselineskip0pt\lineskip.9pt
  \ialign{$\mathsurround=0pt#1\hfil##\hfil$\crcr#2\crcr\sim\crcr}}}
\newcommand {\apgt} {\ {\raise-.5ex\hbox{$\buildrel>\over\sim$}}\ }
\newcommand {\aplt} {\ {\raise-.5ex\hbox{$\buildrel<\over\sim$}}\ }
\newcommand{\bear}{\be\begin{array}}

\newcommand{\vecs}[1]{\mathbf{#1}{\lower-.2em\hbox{}^{2}}}
\newcommand{\veck}[1]{\mathbf{#1}{\kern-.2em\hbox{}^{2}}}
\def\bc{\begin{center}}
\def\ec{\end{center}}

\newcommand{\bea}{\begin{eqnarray}}
\newcommand{\eea}{\end{eqnarray}}

\newcommand{\Li}{\mbox{Li}_2}

\newcommand{\ii}{\mbox{i}}

\newcommand{\be}{\begin{equation}}
\newcommand{\ee}{\end{equation}}
\newcommand{\bg}{\begin{gather}}
\newcommand{\foma}{\end{gather}}

\newcommand{\nn}{\nonumber}

\newcommand{\br} [1]{ \left( #1 \right) }

\def\dd{\mathrm d}

\def\phi{\varphi}

\newcommand\ba{\begin{eqnarray}}
\newcommand\ea{\end{eqnarray}}

\begin{document}
\begin{frontmatter}

\title{Dirac tensor with heavy photon}

\author{V.V.~Bytev, E.A.~Kuraev}
\address{Bogoliubov Laboratory of Theoretical Physics, \\
JINR,\ Dubna, \ 141980 \ \  Russia \\
{\tt e-mail:} kuraev@theor.jinr.ru
}

\author{E.S. Scherbakova}
\address{Hamburg University,\\
Hamburg, \ 22767, \ Germany\\
{\it e-mail:}scherbak@mail.desy.de}

\begin{abstract}
For the large-angles hard photon emission by initial leptons in process
of high energy annihilation of $e^+ e^- \to $ to hadrons
the Dirac tensor is obtained, taking into account the lowest order radiative corrections.
The case of large-angles emission of two hard photons by initial leptons is considered.
This result is being completed by the kinematics case of collinear hard photons emission
as well as soft virtual and real photons and can be used for construction of Monte-Carlo generators.
\end{abstract}


\end{frontmatter}

\section{Introduction}

The problem of precise knowledge of the cross section of annihilation $e^{+} e^{-}$ to hadrons
caused by the long staying problem of theoretical estimation of muon anomalous magnetic moment $g-2$~\cite{g-2}:
\begin{eqnarray}
a_\mu^{\mathrm hadr} &=& \Bigl(\frac{g-2}{2}\Bigr)_{\mu}
= \frac{1}{3}\Bigl(\frac{\alpha}{\pi}\Bigr)^2\int\limits^\infty_{4 m_\pi^2} \frac{\dd s}{s} R(s) K^{(1)}(s),
\\ \nonumber
R(s) &=& \frac{\sigma^{e\bar e\to{\mathrm hadr}}}{\sigma^{e\bar e\to{\mu\bar\mu}}},\quad
K^{(1)}(s) = \int\dd x\frac{x^2 (1-x)}{x^2+\rho(1-x)}, \, \rho = \frac{s}{m_\mu^2}.
\end{eqnarray}
Extraction of cross-section  $e^{+} e^{-} \to \gamma^*\to{\mathrm hadrons}$ from experimental
data is one of the main problem of modern experimental physics. The Monte-Carlo programs creation
which takes into account the emission of real photons
by the initial leptons is the motivation of this paper.

Dirac tensor (cross-symmetry partner of Compton tensor) i.e. the bilinear combination
of the currents of hard photon emission averaged on leptons spin states and summed
on photon polarization states takes the contribution on Born level and
the ones arising from 1-loop correction. Infrared divergences are parametrized
by the introduction "photon mass" $\lambda$. In the final answer it is removed in a usual way
by adding the contribution from additional soft photon emission.

We don't consider photon emission by the final charged particles as well as the effects
of charge-add interference of emission of virtual or real photon emission from leptons and hadrons.
So the Dirac tensor obtained in such way is universal.

The paper is organized as follow. In the part 2 the relation of Dirac tensor with cross section of the radiative
annihilation of lepton pair to hadros is clarified. We put the Born level expression for Dirac tensor and rerive
the general form of radiative correction to it by using the symmetry relation.

In part 3 we obtain the contribution arising from mass operator of positron
and vertex function for the case when positron and photon are on mass shell.
In the part 4 we consider the contribution from vertex function for the case
of electron on mass shell
and the box-type Feynman amplitude with electron, positron and one of photons on mass shell.

In section 5 we analyze the total result for Dirac tensor, adding the emission of additional soft photon
contributions, which provide the infrared divergences free final result.
Limiting case of almost collinear hard photon emission is considered
and some numerical estimation are given.

We put the form of hadronic tensor for several final states:
$\gamma^*\to \pi^{+}\pi^{-}$,  $\mu^{+}\mu^{-}$,  $\rho^{+}\rho^{-}$.
In Appendixes A and B the details of calculation are presented.
In Appendix C the contribution
to Dirac tensor for the case of two hard photon emission is given.

\section{General analysis}

\begin{figure}[h]
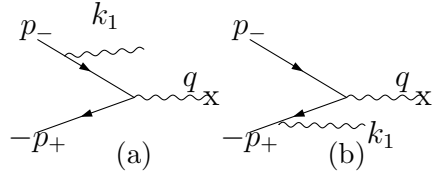

 \begin{center}
\unitlength=1.bp%
\begin{feynartspicture}(160,80)(2,1)
\input born.tex
\end{feynartspicture}
 \caption{
  Diagrams contributing in Born level.
				}
\label{born}
 \end{center}
\end{figure}

The Born level matrix element of hard photon emission by initial leptons in process
of annihilation $e^+$, $e^-$ to hadrons through the single virtual photon intermediate state
\begin{equation}
\label{mainproc}
e^+(p_{+}) + e^-(p_{-})  \rightarrow \gamma^*(q) + \gamma(p_1) \rightarrow \gamma(p_1) + h(q)
\end{equation}
has a form (see Fig.~\ref{born})

\begin{gather}
M = \frac{(4\pi\alpha)^{3/2}}{q^2} \bar{v}(p_{+}) O^{(B)}_\rho u(p_{-}) H_\rho(q),\\ \nonumber
O^{(B)}_\rho = \gamma_\rho\frac{\hat{p}_{-} -\hat{p}_1}{-\chi_{-}} \hat e
 + \hat e \frac{-\hat{p}_{+} +\hat{p}_1}{-\chi_{+}} \gamma_\rho,
\label{LogsDef}
\end{gather}
where $\hat e(p_1)$ is polarization vector of the real photon. The $H_\rho(q)$ is
the current describing the conversation of virtual photon whith momentum $q$
 to hadronic state.
We will restrict ourselves by kinematics conditions of large-angles scattering.
\begin{gather}
s = 2 p_{+} p_{-}, \quad
\chi_\pm = 2 p_1 p_{\pm}, \quad
p_1^2 = 0, \quad
p^2_\pm = m^2, 
\\ \nonumber
s - \chi_{+}  - \chi_{-} = q^2, \quad
q^2 > 0,\quad
s \sim q^2\sim \chi_{+} \sim \chi_{-} \gg m^2. \quad
 \nonumber
\end{gather}
In 
expressions
below we put $m=0$ everywhere except the denominators of loops integrals.

Cross section can be expressed in terms of the summed on spin states
of the module of matrix element square:
\begin{gather}
 \nonumber
\sum_{spin} | M | ^2 = (4 \pi\alpha)^3\frac{4 B_{\rho\rho_1} H_{\rho\rho_1}}{(q^2)^2},
\\
 \nonumber
B_{\rho\rho_1} = \frac{1}{4} \mathrm{Tr}\,\hat p_{+} O_\rho \hat p_{-} \bar O_{\rho_1},\quad
H_{\rho\rho_1} = \sum_{spin} H_\rho(q) H^*_{\rho_1}(q).
\end{gather}
The differential cross-section can be written as:
\begin{gather}
 \nonumber
\dd\sigma^{e^+e^-\to\gamma X} = \frac{1}{8s}\sum_{spin}|M|^2 \frac{\dd^3 p_1}{2\omega(2\pi)^3}\dd\Gamma_f, \\
\dd \Gamma_f = (2\pi)^4\delta^4\Bigr(p_{+}+p_{-}-p_1-\sum_{f} q_i\Bigl)\prod_f\frac{\dd^3q_i}{2\varepsilon_i(2\pi)^3}.
\end{gather}
For differential hard photon cross section we obtain
\begin{gather}
\frac{\omega_1 \, \dd\sigma^{e^+e^-\to\gamma X}}{\dd^3 p_1 \dd \Gamma_f} =
\frac{2\alpha^3}{s (q^2)^2} H_{\rho \rho_1} B_{\rho \rho_1}, \quad
\end{gather}
where
\begin{gather}
B_{\rho \rho_1} = \! B_g \tilde{g}_{\rho \rho_1}
\!\!+\! B_{++}\!\tilde{p}_{+\rho}\tilde{p}_{-\rho_1}
\!\!+\! B_{--}\! \tilde{p}_{-\rho}\tilde{p}_{-\rho_1}
\!\!+ \! B_{+-}\! (\tilde{p_-}\tilde{p_+})_{\rho\rho_1}, \nn \\
(p_+p_-)_{\rho\rho_1}=p_{+\rho}p_{-\rho_1}+p_{+\rho_1}p_{-\rho}.
\end{gather}
The quantities with the "tilde" are defined as
\begin{gather}
\tilde{g}_{\rho \rho_1} = g_{\rho \rho_1} - \frac{1}{q^2} q_\rho q_{\rho_1},\quad
\tilde{p}_{\pm \rho} = p_{\pm\rho} - q_\rho \frac{p_\pm q}{q^2}.
\end{gather}
In the Born approximation (see Fig.~\ref{born}) we have
\begin{gather}
B^B_g =\frac{1}{\chi_{+} \chi_{-}} (2 s q^2 + \chi^2_{+} + \chi^2_{-}), \quad
B^B_{++} = B^B_{--} = \frac{4 q^2}{\chi_{+} \chi_{-}},\quad
B^B_{\pm}  = 0.
\end{gather}
For $q^2=0$ we reproduce the Dirac cross-section of $e^-e^+\to\gamma\gamma$:
\begin{gather}
\frac{\dd\Gamma}{\dd {\mathcal O}_1} = \frac{2\alpha^2}{s}
\frac{\chi_{+}^2 + \chi_{-}^2}{\chi_{+}\chi_{-}}.
\end{gather}
Below we concentrate on calculation of one-loop radiative correction to Dirac tensor.

Let us now show that in considering of the corrections one can restrict only by
half of full set of Feynman diagrams for considering process (2).
So we put $O_\rho = O^-_\rho + O^+_\rho$
separating the contribution of emission from electron leg $ O^-_\rho $ and positron one $O^+_\rho$
(see Fig.~\ref{born} for the Born case and Fig.~\ref{Feydiag} for the 1-loop corrections).

One can show that using the cyclic property of the trace as well as the mirror property
\begin{gather}
\mathrm{Tr}\, \hat a_1 \hat a_2 \cdots \hat a_{2n} =
\mathrm{Tr}\, \hat a_{2n} \cdots \hat a_2 \hat a_1, \nn
\end{gather}
that the total contribution to the Dirac leptonic tensor can be written as:
\begin{eqnarray}
\mathrm{Tr} \, \hat p_{+} O^{(1)}_\rho \hat p_{-}  \bar O^B_{\rho_1}
&+& \mathrm{Tr}\, \hat p_{+} O^{B}_{\rho} \hat p_{-} \bar{O}^{(1)}_{\rho_1}
\nonumber \\
&=&
(1+\Delta_{\rho\rho_1})(1+{\mathcal P}) \mathrm{Tr} \, \hat p_{+} O^+_\rho \hat p_{-} \bar O^B_{\rho_1}.
\end{eqnarray}
Here the exchange operations acting as
\begin{gather}
\Delta_{\rho \rho_1} F_{\rho \rho_1} =F_{\rho_1 \rho},\nonumber\\
{\mathcal P} F(p_{+}, p_{-}, p_1) = F(-p_{-}, -p_{+}, -p_1), \nonumber\\
{\mathcal P} F(s,q^2,\chi_{+}, \chi_{-}) = F(s,q^2,\chi_{-}, \chi_{+})
\equiv \tilde F.
\label{calP}
\end{gather}
Here and below we imply only the real part of leptonic tensor.

\section{ One-loop corrections.Real photon vertex and self energy contribution}

\begin{figure}[h]
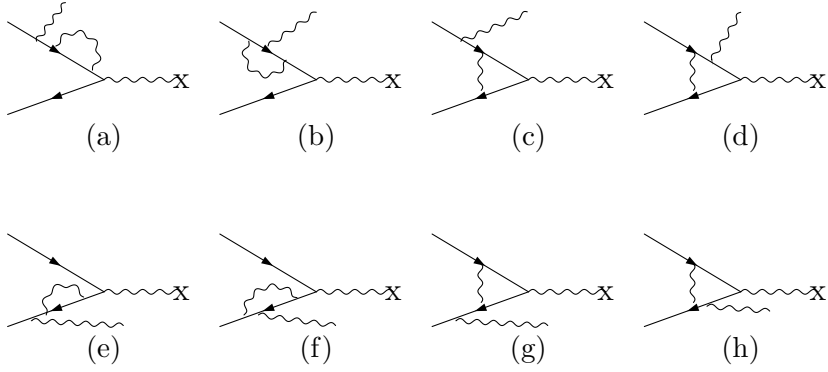

 \begin{center}
\unitlength=1.bp%
\begin{feynartspicture}(320,160)(4,2)
\input diag.tex
\end{feynartspicture}
 \caption{
  Diagrams contributing  in 1-loop level.
				}
\label{Feydiag}
 \end{center}
\end{figure}

The virtual correction of lowest order is described by 8 Feynman diagrams
shown on  the Fig.~\ref{Feydiag}.

Let us distinguish contribution of FD Fig.~(\ref{Feydiag}(e-h)) to 3 classes
\begin{gather}
{\mathrm Tr} \, \hat p_{+} O^+_\rho p_{-}  \bar O^B_{\rho_1}
= T^{\mathrm{box}}_{\rho\rho_1} + T^{\mathrm{verx}}_{\rho\rho_1} +T^{\mathrm{\Sigma}}_{\rho\rho_1}
\end{gather}
with $T^{\mathrm{box}}$ and $T^{\mathrm{verx}}$ correspond to Fig.~(\ref{Feydiag} (h,g))
and $T^{\mathrm{\Sigma}}$  to Fig.~(\ref{Feydiag}~e,f).

Consider first the contribution to matrix element arising from Feynman diagram  Fig.~\ref{Feydiag} e,f.

Matrix element of FD  Fig.~(\ref{Feydiag} (e)) contain the mass operator of electron $\Sigma(\hat p)$.
In kinematics conditions of our problem ($\chi_{+} \gg m^2$) we obtain~\cite{AxBer}:
\begin{gather}
M_e = \frac{\alpha}{2\pi} (\frac{3}{2} + \frac{1}{2} l_{+} - l_\lambda) \bar v(p_{+})\hat e
\bigl(\frac{-\hat p_{+} + \hat p_1}{-\chi_{+}}\bigr)\gamma_\rho u(p_{-}),\\
l_{\pm} =\ln\frac{\chi_{\pm}}{m^2}, \,\, l_\lambda =\ln\frac{m^2}{\lambda^2},
\nonumber
\end{gather}
where $\lambda$ is the so-called "photon mass".

Matrix elements of FD Fig.~(\ref{Feydiag} (f)) contain the vertex function with real photon~\cite{AxBer}.
\begin{gather}
M_f = \frac{\alpha}{4\pi} \bar v(p_{+})\!\!
\int\!\  \frac{\dd^4 k}{i\pi^2}
\frac{ \gamma^\lambda (-\hat p_{+}\!-\!\hat k) \hat \e (-\hat p_{+}+\hat p_1-\hat k) \gamma^\lambda((-\hat p_{+}\! +\! \hat p_1)\gamma_\rho}{(0)(\bar 2)(q)}
\frac{1}{-\chi_{+}}u(p_{-}).
\nonumber
\end{gather}
We use here the notations (see eq. (\ref{design}))

Using the relevant loop integrals, obtained in~\cite{Arbuzov:2010zza} (see Appendix) we have
a matrix elements of FD Fig.~(\ref{Feydiag}(f)) which contain the vertex function with real photon~\cite{AxBer}
\begin{gather}
M_f = \frac{\alpha}{2\pi} \bar v (p_{+}) [ - \frac{1}{\chi_{+}}
( l_{+} -\frac{1}{2}) \hat p_1 \hat e
 + \hat e (l_\lambda -\frac{1}{2} l_{+} -\frac{3}{2})]
\frac{-\hat p_{+} + \hat p_1}{-\chi_{+}} \gamma_\rho u(p_{-}).
\end{gather}

As a result we obtain the infra-red free and gauge-invariant expression:
\begin{gather}
M_e+M_f = \frac{\alpha}{\pi} \Phi_{+} \bar v (p_{+})  \hat p_1 \hat e \gamma_\rho u(p_{-}),\quad
\Phi_{+} = \frac{1}{2 \chi_{+}} ( l_{+} -\frac{1}{2}).
\end{gather}

Inserting this expression to the relevant part of $O^+_\rho$ we obtain for $T^\Sigma_{\rho\rho_1}$
\begin{eqnarray}
\label{last}
T^\Sigma_{\rho\rho_1} &=&- \Phi_{+} \mathrm{Tr} \, \hat p_{+} \hat p_1 \gamma_\lambda\gamma_\rho \hat p_{-}
    \bigl[
\frac{1}{\chi_{-}}\gamma_\lambda (\hat p_{-} -\hat p_1) \gamma_{\rho_1}
		+ \frac{1}{\chi_{+}}\gamma_{\rho_1} (-\hat p_{+} + \hat p_1)\gamma_\lambda\bigr]
,
\end{eqnarray}
where we used relation $p_{1\rho} = (p_{+} + p_{-})_\rho$ keeping mind the gauge invariance of hadronic
tensor $q_\mu H_{\mu\nu}=0$.
Expression~(\ref{last}) leads to the  form for contributions of diagramms  Fig.~(\ref{Feydiag}(e,f)):
\begin{eqnarray}
T^\Sigma_{\rho\rho_1} &=&-4 \Phi_{+} [2 p_{-\rho}p_{-\rho_1}
    (\frac{q^2}{\chi_{-}} -1) + 2 p_{-\rho}p_{+\rho_1} (\frac{s}{\chi_{-}}-1) - (s-\chi_{-})g_{\rho\rho_1}].	
\end{eqnarray}

Applying the operation $1+\Delta_{\rho\rho_1}$ and $1+\mathcal P$ we obtain the full result:
\ba
(1+\Delta_{\rho\rho_1})&\times&(1+\P)T^\Sigma_{\rho\rho_1}
\nn\\
&=&B^\Sigma_g \tilde{g}_{\rho\rho_1}+B^\Sigma_-\tilde{p}_{-\rho}\tilde{p}_{-\rho_1}+
B^\Sigma_+ \tilde{p}_{+\rho}\tilde{p}_{+\rho_1}+B^\Sigma_{+-}(\tilde{p}_+\tilde{p}_-)_{\rho\rho_1},
\ea
with
\begin{gather}
B^\Sigma_g=\frac{4}{\chi_-\chi_+}[s c-\chi_-^2-\chi_+^2]l_s+T^\Sigma_g, \quad
B^\Sigma_-=-\frac{8}{\chi_-\chi_+}[q^2-\chi_-]l_s+T^\Sigma_-, \nn \\
B^\Sigma_+=-\frac{8}{\chi_-\chi_+}[q^2-\chi_+]l_s+T^\Sigma_+, \quad
B^\Sigma_{+-}=-\frac{4}{\chi_-\chi_+}[q^2+s]l_s+T^\Sigma_{+-},
\nn\\
c=\chi_++\chi_-.
\end{gather}
Here we use the notation
\begin{gather}
T^\Sigma_g=-2\frac{s-\chi_-}{\chi_+}[1+2l_{sp}]-2\frac{s-\chi_+}{\chi_-}[1+2l_{sm}], \nn \\
T^\Sigma_-=\frac{4}{\chi_+\chi_-}[q^2-\chi_-][1+2l_{sp}], \quad
T^\Sigma_+=\frac{4}{\chi_+\chi_-}[q^2-\chi_+][1+2l_{sm}]; \nn \\
T^\Sigma_{+-}=\frac{2}{\chi_+\chi_-}[s-\chi_-][1+2l_{sp}]+
\frac{2}{\chi_+\chi_-}[s-\chi_+][1+2l_{sm}],
\nn\\
l_{sp}=l_s-l_+,\quad  l_{sm}=l_s-l_-.
\label{Tsigma}
\end{gather}
\section{Vertex and box type diagram contributions}

Contribution of Fig.~(\ref{Feydiag}(g,h)) can be written as
\begin{gather}
T^{\mathrm box}_{\rho\rho_1} + T^{\mathrm vert}_{\rho\rho_1}
=\frac{S_1}{\chi_{-}}+\frac{S_2}{\chi_{+}}-\frac{C_1}{\chi_{-}\chi_{+}}-\frac{C_2}{\chi_{+}^2},
\end{gather}
with
\begin{eqnarray}
S_1 &=& \int \frac{1}{4} \frac{\dd^4 k}{i\pi^2} \frac
{
 {\mathrm Tr}\, \hat B_\rho \hat p_{-}
\gamma_\eta (\hat p_{-} - \hat p_1) \gamma_{\rho_1}
}{(0)(2)(\bar 2)(q)},
\nn\\ \nn
S_2 &=&  \int \frac{1}{4} \frac{\dd^4 k}{i\pi^2} \frac{
 {\mathrm Tr}\, \hat B_\rho \hat p_{-}
\gamma_{\rho_1} (- \hat p_{+} + \hat p_1) \gamma_\eta
}{(0)(2)(\bar 2)(q)},
\\
\nonumber
C_1 &=& \int \frac{1}{4} \frac{\dd^4 k}{i\pi^2}
 \frac{
 {\mathrm Tr}\, \hat V_\rho \hat p_{-}
\gamma_\eta (\hat p_{-} - \hat p_1) \gamma_{\rho_1}
}{(0)(2)(q)},
\\ \nn
C_2 &=&  \int \frac{1}{4} \frac{\dd^4 k}{i\pi^2}
\frac{
 {\mathrm Tr}\, \hat V_\rho \hat p_{-}
\gamma_{\rho_1} (- \hat p_{+} + \hat p_1) \gamma_\eta
 }{(0)(2)(q)},
\\
\nonumber
\hat B_\rho &=&  \hat p_{+} \gamma_\lambda ( - \hat p_{+} - \hat k)\gamma_\eta
(-\hat p_{+} +\hat p_1 - \hat k)\gamma_\rho ( \hat p_{-}-\hat k) \gamma_\lambda,
\\
\hat V_\rho &=&  \hat p_{+} \gamma_\eta ( - \hat p_{+} + \hat p_1)\gamma_\lambda
(-\hat p_{+} +\hat p_1 - \hat k)\gamma_\rho ( \hat p_{-}-\hat k) \gamma_\lambda.
\end{eqnarray}
Using the loop integrals listed in Appendix A we obtain:
\ba
T^{\mathrm box}_{\rho\rho_1} + T^{\mathrm vert}_{\rho\rho_1}
&=&D_g g_{\rho\rho_1}+D_-p_{-\rho}p_{-\rho_1}+
D_+ p_{+\rho}p_{+\rho_1}
\nn\\
&+&D_{+-}p_{+\rho}p_{-\rho_1}+D_{-+}p_{-\rho}p_{+\rho_1},
\ea
Applying the interchange operator $1+\Delta_{\rho\rho_1}$ and $1+\P$, $\tilde D(\chi_+,\chi_-)=\P D(\chi_-,\chi_+)$,
and rearranging the gauge-invariance we put it in the form:
\begin{gather}
(1+\Delta_{\rho\rho_1})(1+\P)(T^{\mathrm box}_{\rho\rho_1} + T^{\mathrm vert}_{\rho\rho_1})=
B^{VB}_g \tilde{g}_{\rho\rho_1}+B^{VB}_-\tilde{p}_{-\rho}\tilde{p}_{-\rho_1}
\nn\\
+B^{VB}_+ \tilde{p}_{+\rho}\tilde{p}_{+\rho_1}+B^{VB}_{+-}(\tilde{p}_+\tilde{p}_-)_{\rho\rho_1},
\nn
\end{gather}
where
\begin{gather}
B^{VB}_g=2(\tilde D_g+D_g),\quad B^{VB}_-=2(D_-+\tilde D_+),\quad B^{VB}_+=2(D_++\tilde D_-)
\nn\\
B^{VB}_{+-}=D_{+-}+D_{-+}+\tilde D_{+-}+\tilde D_{-+}.
\end{gather}
Here we could see by construction that
\begin{gather}
B^{VB}_g=\tilde B^{VB}_g,\quad B^{VB}_-=\tilde B^{VB}_+,\quad B^{VB}_{+-}=\tilde B^{VB}_{-+},
\end{gather}
and in explicit form $B^{VB}_i$ are:
\begin{gather}
B^{VB}_g=\frac{4s c-8s^2}{\chi_-\chi_+}l_s+\frac{2\chi_-^2+2\chi_+^2-4s c+4s^2}{\chi_-\chi_+}[l_s^2+2(l_s-1)l_\lambda-l_s]+ T^{VB}_g, \nn \\
B^{VB}_-=\frac{8\chi_+-8s}{\chi_-\chi_+}l_s+\frac{8q^2}{\chi_-\chi_+}[l_s^2+2(l_s-1)l_\lambda-l_s]+ T^{VB}_-, \nn \\
B^{VB}_+=\frac{8\chi_--8s}{\chi_-\chi_+}l_s+\frac{8q^2}{\chi_-\chi_+}[l_s^2+2(l_s-1)l_\lambda-l_s]+ T^{VB}_+, \nn \\
B^{VB}_{+-}=\frac{4(s+q^2)}{\chi_-\chi_+}l_s+ T^{VB}_{+-}, \nn
\end{gather}
where the  expressions  $T^{VB}_i$ contains non-leading terms.

These quantities contain the ultra-violet cut off logarithm $L=\ln\frac{\Lambda^2}{m^2}$
which is eliminated by standard regularization procedure~\cite{AxBer}
$L\to 2 l_\lambda -9/2$.

Collecting the leading terms whcih contains the large logarithm $l_s$ and infrared one $l_\lambda$
we obtain:
\begin{gather}
(B_g^{VB}+B_g^\Sigma)_{leading}=2B_g^B(l_s^2+2(l_s-1)L_\lambda-3l_s),
\nn\\
(B_+^{VB}+B_+^\Sigma)_{leading}=(B_-^{VB}+B_-^\Sigma)_{leading}=2B_+^B(ls^2+2(l_s-1)L_\lambda-3l_s),
\nn\\
(B_{+-}^{VB}+B_{+-}^\Sigma)_{leading}=0.
\end{gather}
\section{Discussion, explicit form of tensor structures.}

The infrared divergences constrained in contribution of virtual photon emission canceled
when takes into account the emission of additional soft photon (center-of mass of $e^+ e^-$
initial is implied)
\begin{eqnarray}
\dd \sigma^\gamma_{\mathrm soft} &=& \delta_{\mathrm soft} \dd \sigma_{\mathrm B}, \nn \\
\delta_{\mathrm soft} &=& - \frac{4\pi\alpha}{16\pi^3}\int\frac{\dd^3k}{w}
  ( - \frac{p_{-}}{p_{-} k}+\frac{p_{+}}{p_{+} k})^2, \,\,w<\Delta\varepsilon \ll\sqrt{s}/2,
\end{eqnarray}
where $w = \sqrt{k^2+\lambda^2}$.
Using the standard integrals we obtain
\begin{gather}
\delta_{\mathrm soft} =\frac{\alpha}{\pi} [( l_s-1)(l_\lambda+2\ln\frac{\Delta E}{E})
+ \frac{1}{2}l_s^2 - \frac{\pi^2}{3}].
\end{gather}
Summing all contributions we find Dirac tensor:
\begin{eqnarray}
B_{\rho\rho_1} &=& (B_g^B \tilde{g}_{\rho \rho_1}
\!\!+\! B_{++}^B\!\tilde{p}_{+\rho}\tilde{p}_{-\rho_1}
\!\!+\! B_{--}^B\! \tilde{p}_{-\rho}\tilde{p}_{-\rho_1}
)\nn\\
&\times&( 1 +\frac{\alpha}{\pi}( l_s-1)(\frac{3}{2}+2\ln\frac{\Delta E}{E}) +\frac{\alpha}{\pi}(-\frac{\pi^2}{3}+\frac{3}{2}))\nn \\
&-&\frac{\alpha}{4\pi}[T_g \tilde{g}_{\rho\rho_1}+T_-\tilde{p}_{-\rho}\tilde{p}_{-\rho_1}+
T_+ \tilde{p}_{+\rho}\tilde{p}_{+\rho_1}+T_{+-}(\tilde{p}_+\tilde{p}_-)_{\rho\rho_1}].
\label{eq.final}
\end{eqnarray}
Quantities $T_i=T^\Sigma_i+T^{VB}_i$ are free from infrared singularities and
do not contain large logarithms. Quantities $T^\Sigma_i$ are given in (\ref{Tsigma}). Quantities $T^{VB}_i$ are given in Appendix \ref{non-lead}.

\begin{figure}
\begin{center}
\includegraphics[scale=0.8]{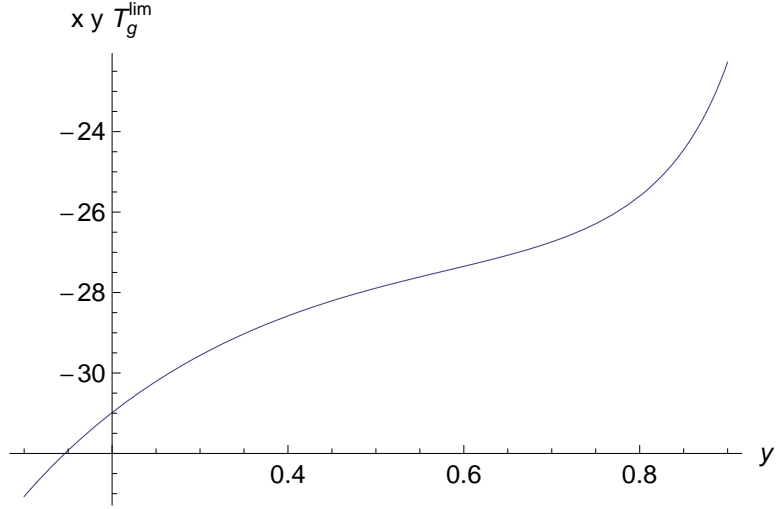}
\caption{ $x yT_g^{lim}$ as a function of $y$ at $x=0.1$}
\label{FigLim}
\end{center}
\end{figure}

Expressions for $T_i$ contains in addition nonphysical singularities $\chi_\pm^{-2}$, $\chi_\pm^{-3}$.
Nevertheless one can be convinced in cancelation of terms proportional  $\chi_\pm^{-3}$ with structure G (see eq. (\ref{strG}))
and terms  $\chi_\pm^2$ in convolution of $T_i$. To be defined let us consider the case of small values of $\chi_-$,
$m^2\ll \chi_-\ll s \sim  q^2\sim \chi_+$. It corresponds to the kinematics $p_1=y p_-$.
In this case non-leading terms containing poles  could be put in the form:
\begin{gather}
 T_g \tilde{g}_{\rho \rho_1}
\!\!+\! (T_-+\bar y^2T_+-2\bar y T_{+-})\!\tilde{p}_{-\rho}\tilde{p}_{-\rho_1},\quad \bar y=1-y, \quad y=\frac{\chi_+}{s}
\end{gather}
and this combination contains only the lowest order pole $\chi_-^{-1}$.
The Dirac tensor in the limit $m^2\ll \chi_- \ll s$ has a form:
\begin{gather}
B^{lim}_{\rho\rho_1}=B^{B,\, lim}_{\rho\rho_1}(1+\frac{\alpha}{\pi}(l_s-1)(\frac{3}{2}+2\ln\frac{\Delta E}{E})+\frac{\alpha}{\pi}(\frac{3}{2}-\frac{\pi^2}{3}))
\nn\\
-\frac{\alpha}{4\pi}T_g^{lim}(\tilde{g}_{\rho \rho_1}+\frac{4\bar y}{s}\tilde{p}_{-\rho}\tilde{p}_{-\rho_1}),
\end{gather}
where
\begin{gather}
B^{B,\, lim}_{\rho\rho_1}=\frac{1+\bar y^2}{xy}(\tilde{g}_{\rho \rho_1}+\frac{4\bar y}{s}\tilde{p}_{-\rho}\tilde{p}_{-\rho_1}),
\quad \bar x=1-x,\quad x=\frac{\chi_-}{s},
\nn\\
T_g^{lim}=\frac{1}{xy}\biggl(16-18y+10y^2+4(1+\bar y^2)\br{\ln\bar y\,\ln \frac{y}{x}-\Li\br{\frac{1}{1-y}}-\frac{\pi^2}{2}}
\nn\\
-8y\bar y \ln y+(-12+20 y-14 y^2)\ln\bar y\biggr),
\end{gather}
Values of $x yT_g^{lim}$ as a function of $y$ and $x=0.1$ are presented at Fig. \ref{FigLim}.

Obtained formula have a power accuracy as well as we systematically omit the terms of order $m^2/s$ compared to ones of order of  unity.

Similar properties have a cross-channel-Compton tensor with one real and another virtual (space-like) photon \cite{KFM87},
where terms of order $m^2/s$ was taken into account.

\section{Samples of hadronic tensors.}

Hadronic tensor is the summed on spin states of bilinear combination of matrix elements $M_\rho M^*_{\rho_1}$,
where the current  $M_\rho$ describes the conversion of heavy time-like photon to some set of hadrons.
For the case of creation of a pair of charged pseudoscalar mesons
($\pi^+\pi^-$, $K^+K^-$, ...) we have
\begin{eqnarray}
H^{p_+^s p_{-}^s}_{\rho\rho_1} &=& (p_{+}^s -p_{-}^s)_\rho (p_{+}^s - p_{-}^s)_{\rho_1}, \,\, q = p_{+}^s + p_{-}^s.
\nonumber
\end{eqnarray}
For conversion to pair of charged spin 1/2 fermions
$\gamma \to \mu^+ (p_{+}) + \mu^-(p_{-})$ we have

\begin{eqnarray}
 H^{\mu_+ \mu_{-}}_{\rho\rho_1} = 4 [ p_{+\rho}^m p_{-\rho_1}^m + p_{+\rho_1}^m p_{-\rho}^m
                 - \frac{q
                 ^2}{2} g_{\rho\rho_1}].
\end{eqnarray}
For creation of a pair of charged vectors mesons $\rho^+\rho^-$, $K^{*+}K^{*-}$
one obtaines
\begin{eqnarray}
H^{q_+ q_{-}}_{\rho\rho_1} &\approx& q^2 (8-2\eta)(g_{\rho\rho_1} - \frac{q_\rho q_{\rho_1}}{q^2})
\\ \nonumber
   & +& (q_{+} - q_{-})_\rho (q_{+} - q_{-})_{\rho_1} ( 3-5\eta+\frac{9}{4}\eta^2),\\
&&\eta =\frac{q^2}{m^2_\rho}, \quad q = q_{\rho +} + q_{\rho -}.
\end{eqnarray}

The gauge invariance requirement $H_{\rho\rho_1} q_\rho = H_{\rho\rho_1} q_\rho = 0$ is
fulfilled.

\subsection*{Acknowledgements}

One of us (E.A.~Kuraev) is grateful to Hamburg University, II Institute for Theoretical Physics
for warm hospitality in November 2011 where the most part of this work was done.
We grateful to Yu.M.~Bystritskiy for help and to V.~Tayursky and V.~Druzhinin for interest to this problem.
As well he is grateful to RFBR 11-02-00112,  Belorussian 2011 and Heisenberg-Landau 2011  grants for financial support.

\appendix

\section{One-loop Feynman integrals}
\label{App:integrals}

In this section we perform the result of calculation of 4-fold integrals, associated with one-loop Feynman diagram.
Here and below we imply only real part of integrals.
The denominators of integrals defined as
\begin{eqnarray}
(0)&=&k^2-\lambda^2,\nn \\
(2)&=&(p_--k)^2-m^2+\ii0=k^2-2p_-k+\ii0,\nn\\
(\bar{2})&=&(-p_+-k)^2-m^2+\ii0,\nn \\
(q)&=&(p_1-p_+-k)^2-m^2+\ii0.
\label{design}
\end{eqnarray}

The four denominator scalar integral
\begin{gather}
I_{02\bar{2}q}=\int \dd k \frac{1}{(0)(2)(\bar{2})(q)}, \quad \dd k=\frac{\dd^4k}{\ii\pi^2}
\end{gather}
has the form
\begin{gather}
I_{02\bar{2}q}=\frac{1}{s\chi_+}\Big[l_q^2-2l_+l_s-l_sl_l+
2\Li\biggl({1-\frac{q^{2}}{s}}\biggr) - \frac{5 \pi^2}{6}
\Big],
\end{gather}
where the logarithms was denoted in~(\ref{LogsDef}) and
\begin{gather}
l_q=\ln\frac{q^2}{m^2},\quad l_s=\ln\frac{s}{m^2}.
\end{gather}

For the tree and two denominator scalar integrals
\begin{gather}
I_{ijk}=\int \dd k \frac{1}{r}
\end{gather}
with $r=(ij),(ijk),(ijkl)$ where
$i,j,k,l=(0),(2),(\bar{2}),(q)$,
we have following expressions
\begin{align}
I_{0\bar{2}q}&=-\frac{1}{2\chi_+}\Big[l_+^2+\frac{2\pi^2}{3}\Big],
\qquad  I_{02\bar{2}}=\frac{1}{2s}\Big[l_s^2+2l_sl_l-
\frac{4\pi^2}{3}
\Big],& \nn \\
 I_{2\bar{2}q}&=-\frac{1}{2(s-q^{2})}\Big[l_q^2-l_s^2\Big],&\nn\\
I_{02q}&=\frac{1}{\chi_+ + q^{2}}\Big[l_q(l_q-l_+)+
\frac{1}{2}(l_q-l_+)^2+2\Li\biggl({1+\frac{\chi_+}{q^{2}}}\biggr)-\frac{3\pi^2}{2}\Big].&
\end{align}
Two denominator scalar integrals are
\begin{align}
I_{02}&=L+1,& I_{2q}&=L-l_q+1,& I_{0q}&=L-l_++1, \nn\\
I_{0\bar{2}}&=L+1,& I_{2\bar{2}}&=L-L_s+1,& I_{\bar{2}q}&=L-1.\nn
\end{align}

The vector integrals can be defined as
\begin{equation} \label{eq:vecint}
I_{r}^{\mu}=\int\frac{\dd^4k k^\mu}{r}=
a^+_rq_+^\mu+a^-_rq_-^\mu+a^1_rp_1^\mu.
\end{equation}

For the vector integrals with two denominators we have (imaginary part is neglected):
\begin{align}
 a^-_{2q}&=  a^1_{2q}=-  a^+_{2q}=\frac{1}{2}\Big(L-l_q+\frac{1}{2}\Big),&
&a^1_{0q}=-a^+_{0q}=\frac{1}{2}\Big(L-l_++\frac{1}{2}\Big),&\nn \\
 a^-_{2\bar{2}}&=- a^+_{2\bar{2}}=\frac{1}{2}\Big(L-l_s+\frac{1}{2}\Big),&
&a^1_{\bar{2}q}=-\frac{1}{2}a^+_{\bar{2}q}=\frac{1}{2}\Big(L-\frac{3}{2}\Big),&\nn \\
a^-_{02}&=\frac{1}{2}L-\frac{1}{4},&
&a^+_{0\bar{2}}=-\frac{1}{2}L+\frac{1}{4}&
\end{align}
and the coefficients for the vector integrals with three
denominators are
\begin{align}
a^-_{02q}&=\frac{1}{a}\Big(\chi_+I_{02q}+\frac{2\chi_+}{a}l_+
+\frac{q^{2}-\chi_+}{a}l_q\Big),&
&a^+_{02q}=-a^1_{02q}=\frac{1}{a}\Big(l_+-l_q\Big),& \nn\\
a^1_{0\bar{2}q}&=\frac{1}{\chi_+}\Big(-l_++2\Big),&
&a=\chi_++q^{2},& \nn\\
a^+_{0\bar{2}q}&=-I_{0\bar{2}q}-\frac{1}{\chi_+}l_+,&
&a^-_{02\bar{2}}=-a^+_{02\bar{2}}=\frac{1}{s}l_s,& \nn\\
a^-_{2\bar{2}q}&=\frac{1}{c}\Big(l_s-l_q\Big),&
&a^+_{2\bar{2}q}=-I_{2\bar{2}q}+\frac{1}{c}\Big(l_s-l_q\Big),& \nn\\
a^1_{2\bar{2}q}&=\frac{s}{c}I_{2\bar{2}q}+\frac{1}{c}\Big(-l_q+2\Big)
-\frac{2s}{c^2}\Big(l_s-l_q\Big),&
&c=s-q^{2}=\chi_++\chi_-.&
\end{align}
Finally, the coefficient of the vector integral with 4
denominators has the form
\begin{align}
&a^1=\frac{s}{d}\Big(\chi_+A+\chi_-B-sC\Big),&
&a^+=\frac{\chi_-}{d}\Big(\chi_+A-\chi_-B+sC\Big)& \nn\\
&a^-=\frac{\chi_+}{d}\Big(-\chi_+A+\chi_-B+sC\Big),&
&d=2s\chi_+\chi_-,& \nn\\ &A=I_{2\bar{2}q}-I_{0\bar{2}q},&
&B=I_{02q}-I_{2\bar{2}q},& \nn\\
&C=I_{02q}-I_{02\bar{2}}-\chi_+I_{02\bar{2}q}.& &&
\end{align}

The second rank tensor integrals can be parameterized in the form
\begin{multline}
I_{r}^{\mu\nu}=\int\dd k \frac{k_\mu
k_\nu}{r}=\Big[a^g_r g+a^{11}_r
p_1p_1+a^{++}_rq_+q_++a^{--}_rq_-q_-+
a^{1+}_r(p_1q_++q_+p_1)\\
+a^{1-}_r(p_1q_-+q_-p_1)+a^{+-}_r(q_+q_-+q_-q_+)\Big]_{\mu\nu}.
\end{multline}
The coefficients for tensor integral with four denominators are
(we suppressed the index $02\bar{2}q$)
\begin{align}
&a^{1+}=\frac{1}{\chi_+}\Big(A_6+A_7-A_{10}\Big),&
&a^{+-}=\frac{1}{s}\Big(A_2+A_6-A_{10}\Big),&\nn\\
&a^{1-}=\frac{1}{\chi_-}\Big(A_2+A_7-A_{10}\Big),&
&a^{11}=\frac{1}{\chi_-}\Big(A_1-sa^{1+}\Big),& \nn\\
&a^{--}=\frac{1}{s}\Big(A_5-\chi_+a^{1-}\Big),&
&a^{++}=\frac{1}{s}\Big(A_3-\chi_-a^{1+}\Big),& \nn\\
&a^g=\frac{1}{2}\Big(A_{10}-A_2-\chi_+a^{1+}\Big),& &&
\end{align}
with
\begin{align}
&A_1=a^1_{2\bar{2}q}-a^1_{0\bar{2}q},&
&A_6=a^+_{02q}-a^+_{2\bar{2}q},& \nn\\
&A_2=a^-_{2\bar{2}q},&
&A_{7}=a^1_{02q}-\chi_+a^1,& \nn\\
&A_3=a^+_{2\bar{2}q}-a^+_{0\bar{2}q},&
&A_{8}=a^-_{02q}-a^-_{02\bar{2}}-\chi_+a^-,& \nn\\
&A_4=a^1_{02q}-a^1_{2\bar{2}q},&
&A_{9}=a^+_{02q}-a^+_{02\bar{2}}-\chi_+a^+,& \nn\\
&A_5=a^-_{02q}-a^-_{2\bar{2}q},& &A_{10}=I_{2\bar{2}q}.&
\label{eq:ejka}
\end{align}

For the tensor integrals with three denominators $I^{\mu\nu}_{02q}$ we have coefficients\footnote{Formula (2.15) form \cite{Arbuzov:2010zza}
contains a misprint}
\begin{align}
&a^g_{02q}=\frac{1}{4}L+\frac{3}{8}-\frac{q^{2}}{4a}l_q-
\frac{\chi_+}{4a}l_+,& \nn\\ &a^{+-}_{02q}=-a^{1-}_{02q}=\frac{1}{2a}
\Big[\frac{\chi_+}{a}(l_+-l_q)-1\Big],& \nn\\
&a^{++}_{02q}=a^{11}_{02q}=-a^{1+}_{02q}=\frac{1}{2a}(l_q-l_+),&
\nn\\
&a^{--}_{02q}=\frac{1}{a^2}\Big[\chi_+^2I_{02q}+\frac{3\chi_+^2}{a}l_+
-\frac{(q^{2})^2+4 q^{2}\chi_+-3\chi^2_+}{2a}l_q
-\frac{q^{2}+3\chi_+}{2}\Big].&
\end{align}
The coefficients entering into the tensor integral $I^{\mu\nu}_{02\bar{2}}$
are
\begin{gather}
a^g_{02\bar{2}}=\frac{1}{4}(L-l_s)+\frac{3}{8},\quad
a^{++}_{02\bar{2}}=a^{--}_{02\bar{2}}=\frac{1}{2s}(l_s-1),\quad
a^{+-}_{02\bar{2}}=-\frac{1}{2s},
\end{gather}
and the coefficients for the tensor integral $I^{\mu\nu}_{0\bar{2}q}$
are
\begin{align}
&a^g_{0\bar{2}q}=\frac{1}{4}(L-l_+)+\frac{3}{8},&
&a^{1+}_{0\bar{2}q}=\frac{1}{\chi_+}\Big(l_+-\frac{5}{2}\Big),&\nn\\
&a^{11}_{0\bar{2}q}=\frac{1}{2\chi_+}(-l_++2),&
&a^{++}_{0\bar{2}q}=I_{0\bar{2}q}+\frac{1}{2\chi_+}(3l_+-1).
\end{align}
In the case of the tensor integral $I^{\mu\nu}_{2\bar{2}q}$ they have
the form
\begin{align} \label{eq:a22q}
&a^g_{2\bar{2}q}=\frac{1}{2}\Big[\frac{1}{2}L+\frac{3}{4}-
\frac{s}{2c}l_s+\frac{q^{2}}{2c}l_q \Big],\qquad
a^{--}_{2\bar{2}q}=-\frac{1}{2c}(l_q-l_s),&\\
&a^{++}_{2\bar{2}q}=I_{2\bar{2}q}+\frac{3}{2c}(l_q-l_s),\quad\:\:\:\,
\qquad\qquad a^{+-}_{2\bar{2}q}=\frac{1}{2c}(l_q-l_s),& \nn\\
&a^{1-}_{2\bar{2}q}=\frac{1}{c}\Big[-\frac{1}{2}+\frac{s}{2c}l_s-
\frac{s}{2c}l_q\Big],& \\
&a^{1+}_{2\bar{2}q}=\frac{1}{c}\Big[-\frac{5}{2}-sI_{2\bar{2}q}+
\frac{5s}{2c}l_s-\frac{2 q^{2}+3s}{2c}l_q\Big],& \nn\\
&a^{11}_{2\bar{2}q}=\frac{1}{c^2}\Big[4s-q^{2}+
s^2I_{2\bar{2}q}-\frac{3s^2}{c}l_s+
\frac{3s^2-(q^{2})^2+4s q^{2}}{2c}l_q\Big].&
\end{align}

\section{Explicit form of coefficients of non-leading tensor structures}
\label{non-lead}

\begin{align}
    T_g^{VB} &= \br{1+\P}
    \left[
        a_0
        +
         a_1 l_{sq}
        +
        a_2 l_{qp}
        +
        a_4 l_{sp}
        +
        a_6 l_{sq} l_{sp}
        +
        a_8 l_{sq}^2
    \right.
    \nn\\
    &\qquad\qquad\quad\left.
        +
        a_9 \Li\br{1-\frac{q^2}{s}}
        +
        a_{10} \Li\br{1+\frac{\chi_-}{q^2}}
        -
        4\frac{sq^2}{\chi_-^2}G
    \right],
    \label{eq.Tg}
\end{align}
where
\begin{align}
    a_0 &=
    \frac{\pi^2}{3}
    \left[
        -10\frac{\chi_+}{\chi_-}
        +22\frac{s}{\chi_-}
        -4\frac{s^2}{\chi_+\chi_-}
    \right]
    +8\frac{\chi_+}{\chi_-}
    -16\frac{s}{\chi_-}
    +8\frac{s^2}{\chi_+\chi_-},
    \nn\\
    a_1 &=
    \frac{4s}{c}
    +
    10 \frac{\chi_+}{\chi_-}
    -
    20 \frac{s}{\chi_-}
    +
    \frac{8s^2}{\chi_+\chi_-},
    \nn\\
    a_2 &=
    \frac{6\br{s-\chi_+}}{q^2+\chi_+},
    \nn\\
    a_4 &=
    -\frac{4\chi_-}{\chi_+}
    -\frac{4\chi_+}{\chi_-}
    +\frac{4s}{\chi_+}
    +\frac{4s}{\chi_-},
    \nn\\
    a_6 &=
    -\frac{4\chi_-}{\chi_+}
    +\frac{8s}{\chi_+}
    +\frac{4s}{\chi_-}
    -\frac{8s^2}{\chi_+\chi_-},
    \nn\\
    a_8 &=
    \frac{2\chi_+}{\chi_-}
    -\frac{6s}{\chi_-}
    +\frac{4s^2}{\chi_+\chi_-},
    \nn\\
    a_9 &=
    -4\frac{\chi_+}{\chi_-}
    +\frac{12 s}{\chi_-}
    -\frac{8s^2}{\chi_+\chi_-},
    \nn\\
    a_{10} &=
    4\frac{\chi_+}{\chi_-}
    -\frac{4s}{\chi_+}
    -\frac{8s}{\chi_-}
    +\frac{8s^2}{\chi_+\chi_-},
    \nn\\
    G &=
    \Li\br{1-\frac{q^2}{s}}
    +
    \Li\br{1+\frac{\chi_+}{q^2}}
    +l_{sq} l_{sp}-\frac{1}{2}l_{sq}^2+\frac{\pi^2}{6}.
\label{strG}
\end{align}

Note that in the limit $\chi_-\to 0$ quantity G turns to zero.

\begin{align}
    T_+^{VB} &= \P T_-^{VB}, \nn\\
    T_-^{VB} &=
        b_0
        +
        b_1 l_{sq}
        +
        b_2 l_{qp}
        +
        b_3 l_{sm}
        +
        b_4 l_{sp}
        +
        b_5 l_{sq} l_{sm}
        +
        b_6 l_{sq} l_{sp}
        +
        b_7 l_{sq}^2
    \nn\\
    &+
        b_8 \Li\br{1-\frac{q^2}{s}}
        +
        b_9 \Li\br{1+\frac{\chi_-}{q^2}}
        +
        b_{10} \Li\br{1+\frac{\chi_+}{q^2}}
       \nn\\
    &
        -8\frac{(s-\chi_+)^3}{\chi_+\chi_-^3}G+8\frac{s^2}{\chi_+^3}\br{1-\frac{s}{\chi_-}}\P G,
        \label{eq.Tm}
\end{align}
where
\begin{align}
    b_0 &=
    \frac{16}{c}
    \br{1 - \frac{s}{\chi_+} + \frac{s^2}{\chi_+^2}}
    -\frac{32}{\chi_+}
    -\frac{48}{\chi_-}
    +\frac{60s}{\chi_+\chi_-}
    -\frac{16s^2}{\chi_+^2\chi_-}
    \nn\\
    &+
    \frac{\pi^2}{3}
    \left[
         \frac{44}{\chi_+}
        +\frac{44}{\chi_-}
        +\frac{4\chi_+}{\chi_-^2}
        -44\frac{s}{\chi_+\chi_-}
        -8\frac{s}{\chi_-^2}
        +4\frac{s^2}{\chi_+^2\chi_-}
        +4\frac{s^2}{\chi_+\chi_-^2}
    \right]
    \nn\\
    &+
    \frac{1}{\chi_++q^2}
    \br{
        -8 + \frac{4\chi_+}{\chi_-} + \frac{4s}{\chi_+}
    },
    \nn\\
    b_1 &=
    \frac{16s}{c^2}
    \br{-1 + \frac{s}{\chi_+} - \frac{s^2}{\chi_+^2}}
    +
    \frac{16}{c}
    \br{1 - \frac{s}{\chi_+} + \frac{2s^2}{\chi_+^2} + \frac{s^3}{\chi_+^2\chi_-}}
    \nn\\
    &
    -\frac{20}{\chi_+}
    -\frac{32}{\chi_-}
    -\frac{32\chi_+}{\chi_-^2}
    +\frac{8s}{\chi_+^2}
    +\frac{64s}{\chi_+\chi_-}
    +\frac{48s}{\chi_-^2}
    -\frac{48s^2}{\chi_+^2\chi_-}
    -\frac{32s^2}{\chi_+\chi_-^2}
    +\frac{8s^3}{\chi_+^2\chi_-^2},
    \nn\\
    b_2 &=
    \frac{1}{\br{q^2+\chi_+}^2}
    \br{ -8\chi_+ + 4\frac{\chi_+^2}{\chi_-} + 4s }
    +
    \frac{1}{q^2+\chi_+}
    \br{ 4 + 4\frac{\chi_+}{\chi_-} - \frac{8\chi_+^2}{\chi_-^2} - \frac{4s}{\chi_+} },
    \nn\\
    b_3 &=
    -\frac{4}{\chi_+}
    -\frac{4}{\chi_-}
    -\frac{8s}{\chi_+^2}
    -\frac{4s}{\chi_+\chi_-}
    +\frac{8s^2}{\chi_+^2\chi_-},
    \nn\\
    b_4 &=
    \frac{16}{\chi_+}
    +\frac{12}{\chi_-}
    +\frac{24\chi_+}{\chi_-^2}
    -\frac{12s}{\chi_+\chi_-}
    -\frac{24s}{\chi_-^2}
    +\frac{8s^2}{\chi_+\chi_-^2},
    \nn\\
    b_5 &=
    \frac{8}{\chi_+}
    +\frac{8}{\chi_-}
    -\frac{8s}{\chi_+\chi_-}
    +\frac{8s^2}{\chi_+^2\chi_-},
    \nn\\
    b_6 &=
    \frac{16}{\chi_+}
    +\frac{16}{\chi_-}
    +\frac{8\chi_+}{\chi_-^2}
    -\frac{16s}{\chi_+\chi_-}
    -\frac{16s}{\chi_-^2}
    +\frac{8s^2}{\chi_+\chi_-^2},
    \nn\\
    b_7 &=
    -\frac{12}{\chi_+}
    -\frac{12}{\chi_-}
    -\frac{4\chi_+}{\chi_-^2}
    +\frac{12s}{\chi_+\chi_-}
    +\frac{8s}{\chi_-^2}
    -\frac{4s^2}{\chi_+^2\chi_-}
    -\frac{4s^2}{\chi_+\chi_-^2},
    \nn\\
    b_8 &=
    \frac{24}{\chi_+}
    +\frac{24}{\chi_-}
    +\frac{8}{\chi_+\chi_-^2}
    -\frac{24s}{\chi_+\chi_-}
    -\frac{16s}{\chi_-^2}
    +\frac{8s^2}{\chi_+^2\chi_-}
    +\frac{8s^2}{\chi_+\chi_-^2},
    \nn\\
    b_9 &=
    -\frac{8}{\chi_+}
    -\frac{8}{\chi_-}
    +\frac{8s}{\chi_+\chi_-}
    -\frac{8s^2}{\chi_+^2\chi_-},
    \nn\\
    b_{10} &=
    -\frac{16}{\chi_+}
    -\frac{16}{\chi_-}
    -\frac{8\chi_+}{\chi_-^2}
    +\frac{16s}{\chi_+\chi_-}
    +\frac{16s}{\chi_-^2}
    -\frac{8s^2}{\chi_+\chi_-^2}.
\end{align}

\begin{align}
    T_{+-}^{VB} &=(1+\P)\biggl(
        c_0
        +
        c_1 l_{sq}
        +
        c_3 l_{qm}
        +
        c_4 l_{sm}
        +
        c_6 l_{sq} l_{sm}
        +
        c_8 l_{sq}^2
    \nn\\
    &+
        c_9 \Li\br{1-\frac{q^2}{s}}
        +
        c_{10} \Li\br{1+\frac{\chi_-}{q^2}}
        -\frac{8s(s-\chi_+)^2}{\chi_+\chi_-^3}G\biggr),
        \label{eq.Tpm}
\end{align}
where
\begin{align}
    c_0 &=
    \frac{8}{c}
    \br{1 - \frac{s^2}{\chi_+\chi_-}}
    -\frac{4}{q^2+\chi_-}
    -\frac{6}{\chi_-}
    +\frac{10s}{\chi_+\chi_-}
    \nn\\
    &+
    \frac{\pi^2}{3}
    \left[
        -\frac{4}{\chi_-}
        -\frac{4\chi_+}{\chi_-^2}
        +\frac{4s^2}{\chi_+\chi_-^2}
    \right],
    \nn\\
    c_1 &=
    \frac{-8s}{c^2}
    \br{1 - \frac{s^2}{\chi_+\chi_-}}
    +
    \frac{1}{c}
    \br{8 - \frac{12s^2}{\chi_+\chi_-} }
    \nn\\
    &
    +\frac{8}{\chi_-}
    -\frac{8\chi_+}{\chi_-^2}
    +\frac{16s}{\chi_+\chi_-}
    +\frac{24s}{\chi_-^2}
    -\frac{24s^2}{\chi_+\chi_-^2}
    +\frac{4s^3}{\chi_+^2\chi_-^2},
    \nn\\
    c_3 &=
    \frac{-4\chi_-}{\br{q^2+\chi_-}^2}
    +
    \frac{12\chi_-}{\chi_+\br{q^2+\chi_-}},
    \nn\\
    c_4 &=
    \frac{8\chi_-}{\chi_+^2}
    -\frac{8}{\chi_+}
    +\frac{4}{\chi_-}
    -\frac{16s}{\chi_+^2}
    -\frac{8s}{\chi_+\chi_-}
    +\frac{8s^2}{\chi_+^2\chi_-},
    \nn\\
    c_6 &=
    -\frac{8\chi_-}{\chi_+^2}
    -\frac{8}{\chi_+}
    +\frac{8s^2}{\chi_+^2\chi_-},
    \nn\\
    c_8 &=
    \frac{4}{\chi_-}
    +\frac{4\chi_+}{\chi_-^2}
    -\frac{4s^2}{\chi_+\chi_-^2},
    \nn\\
    c_9 &=
    -\frac{8}{\chi_-}
    -\frac{8\chi_+}{\chi_-^2}
    +\frac{8s^2}{\chi_+\chi_-^2},
    \nn\\
    c_{10} &=
    \frac{8\chi_-}{\chi_+^2}
    +\frac{8}{\chi_+}
    -\frac{8s^2}{\chi_+^2\chi_-}.
    \nn
\end{align}


\section{Two hard photon large-angles emission by the initial leptons}

Cross section of 2 photon emission by the initial leptons masses
\begin{eqnarray}
\e^+(p_{+}) + \e^-(p_{-}) \to \gamma(p_1) + \gamma(p_2) + \mathrm{hadr}(q)
\end{eqnarray}
has a form
\begin{eqnarray}
\frac{\dd\sigma^{2\gamma}}{\dd\Gamma_h} &=&\frac{1}{2!}\frac{\alpha^4}{2\pi^2 s}
\frac{H_{\rho\rho_1} O^{(2)}_{\rho\rho_1}}{(q^2)^2} \frac{\dd^2 p_1}{\omega_1} \frac{\dd^2 p_2}{\omega_2},
\quad \omega_1,\omega_2 < \Delta\varepsilon,
\end{eqnarray}
where the factor $1/2!$ takes into account the identity of final-state hard
photons. The relevant contribution to lepton tensor is
\begin{eqnarray}
Q^{(2)}_{\rho\rho_1} &=&\frac{1}{4}\mathrm{Tr}\,p_{+} O^{\sigma\eta}_{12\rho} p_{-}\bar O^{\sigma\eta}_{12\rho};
\\
O^{\sigma\eta}_{12\rho} &=&\gamma_\rho \frac{ \hat p_{-} - \hat p_1 - \hat p_2}{d_{-12} }
\Bigl(\gamma^\eta \frac{ \hat p_{-} - \hat p_1}{d_{-1}}\gamma^\sigma
    + \gamma^\sigma \frac{ \hat p_{-} - \hat p_2}{d_{-2}}\gamma^\eta \Bigr)
	\\ \nonumber
 &+& 	\Bigl(\gamma^\eta \frac{ -\hat p_{+} + \hat p_2}{d_{+2}}\gamma^\sigma
    + \gamma^\sigma \frac{ -\hat p_{+} + \hat p_1}{d_{+1}}\gamma^\eta \Bigr)
 \frac{ - \hat p_{+} + \hat p_1 + \hat p_2}{d_{+12} }\gamma_\rho
	\\ \nonumber
&+& \frac{1}{d_{-1}d_{+2}}\gamma^\sigma (-\hat p_{+} + \hat p_2)\gamma_\rho
(\hat p_{-} - \hat p_1)\gamma^\eta
	\\ \nonumber
&+& \frac{1}{d_{-2}d_{+1}}\gamma^\eta (-\hat p_{+} + \hat p_1)\gamma_\rho
(\hat p_{-} - \hat p_2)\gamma^\sigma,
\end{eqnarray}
and
\begin{eqnarray}
d_{-12} &=& (p_{-} - p_1 - p_2 )^2 - m^2;
\\ \nonumber
d_{-1} &=& (p_{-} - p_1)^2 - m^2; \quad d_{-2} = (p_{-} - p_2)^2 - m^2;
\\ \nonumber
d_{+12} &=& (- p_{+} + p_1 + p_2 )^2 - m^2;
\\ \nonumber
d_{+1} &=& (- p_{+} + p_1)^2 - m^2; \quad d_{+2} = ( p_{+} + p_2)^2 - m^2.
\end{eqnarray}

Tensor $Q^{(2)}_{\rho\rho_1}$ obey the gauge invariance
$Q^{(2)}_{\rho\rho_1} q_\rho = Q^{(2)}_{\rho\rho_1} q_{\rho_1} = 0$ and can be put on the form
\begin{eqnarray}
Q^{(2)}_{\rho\rho_1} &=& A_g \tilde g_{\rho\rho_1}
  + [A_{-} \tilde p_{-} \tilde p_{-}  + A_{+} \tilde p_{+} \tilde p_{+}
  +  A_{11} \tilde k_{1} \tilde k_{1}  + A_{+-} (\tilde p_{+} \tilde p_{-} + \tilde p_{-} \tilde p_{+})
\\ \nonumber
&& \qquad\qquad  +  A_{+1} (\tilde p_{+} \tilde p_{1}+\tilde p_{1} \tilde p_{+})
  + A_{-1} ( \tilde p_{-} \tilde p_{1}+ \tilde p_{1} \tilde p_{-}) ]_{\rho\rho_1},
\end{eqnarray}
coefficients $A_i$ can be obtained in the standard way: constructing the values
\begin{eqnarray}
\nonumber
 B_g, \, B_{11}, \, B_{++}, \, B_{--}, \, ...
 & =& Q_{\rho\rho_1} [g_{\rho\rho_1}, \, p_{1\rho}p_{1\rho_1}, \, p_{+\rho}p_{+\rho_1}, \, ...]
\end{eqnarray}
and solving the set of 7 linear equations.

\end{document}

%% file: born.tex
\FADiagram{(a)}
\FAProp(0.,3.4)(10.,7.)(0.,){/Straight}{-1}
\FALabel(0.,3.93)[t]{$-p_{+}$}
\FAProp(0.,13.)(10.,7.)(0.,){/Straight}{1}
\FALabel(0.,15.)[t]{$p_{-}$}
\FAProp(10.,7.)(18.,7.)(0.,){/Sine}{0}
\FALabel(18.,7.6)[t]{$\mathrm x$}
\FALabel(15.8,9.6)[t]{$q$}
\FAProp(2.8,11.3)(11.,12.)(0.,){/Sine}{0}
\FALabel(7.,16.93)[t]{$k_1$}

\FADiagram{(b)}
\FAProp(0.,3.4)(10.,7.)(0.,){/Straight}{-1}
\FALabel(0.,3.93)[t]{$-p_{+}$}
\FAProp(0.,13.)(10.,7.)(0.,){/Straight}{1}
\FALabel(0.,15.)[t]{$p_{-}$}
\FAProp(10.,7.)(18.,7.)(0.,){/Sine}{0}
\FALabel(18.,7.6)[t]{$\mathrm x$}
\FALabel(15.8,9.6)[t]{$q$}
\FAProp(3.,4.2)(12.,4.)(0.,){/Sine}{0}
\FALabel(13.6,4.5)[t]{$k_1$}

%% file: diag.tex
\FADiagram{(a)}
\FAProp(0.,3.4)(10.,7.)(0.,){/Straight}{-1}
\FAProp(0.,13.)(10.,7.)(0.,){/Straight}{1}
\FAProp(10.,7.)(18.,7.)(0.,){/Sine}{0}
\FALabel(18.,7.6)[t]{$\mathrm x$}
\FAProp(3.,11.)(6.,15.)(0.,){/Sine}{0}
\FAProp(5.,10.5)(8.8,8.)(-1.2,){/Sine}{0}

\FADiagram{(b)}
\FAProp(0.,3.4)(10.,7.)(0.,){/Straight}{-1}
\FAProp(0.,13.)(10.,7.)(0.,){/Straight}{1}
\FAProp(10.,7.)(18.,7.)(0.,){/Sine}{0}
\FALabel(18.,7.6)[t]{$\mathrm x$}
\FAProp(5.,10.5)(10.,14.)(0.,){/Sine}{0}
\FAProp(3.,11.)(7.,9.1)(1.2,){/Sine}{0}

\FADiagram{(c)}
\FAProp(0.,3.4)(10.,7.)(0.,){/Straight}{-1}
\FAProp(0.,13.)(10.,7.)(0.,){/Straight}{1}
\FAProp(10.,7.)(18.,7.)(0.,){/Sine}{0}
\FALabel(18.,7.6)[t]{$\mathrm x$}
\FAProp(3.,11.)(10.,14.)(0.,){/Sine}{0}
\FAProp(5.,10.)(5.,5.8)(0.,){/Sine}{0}

\FADiagram{(d)}
\FAProp(0.,3.4)(10.,7.)(0.,){/Straight}{-1}
\FAProp(0.,13.)(10.,7.)(0.,){/Straight}{1}
\FAProp(10.,7.)(18.,7.)(0.,){/Sine}{0}
\FALabel(18.,7.6)[t]{$\mathrm x$}
\FAProp(7.,9.1)(10.,14.)(0.,){/Sine}{0}
\FAProp(5.,10.)(5.,5.8)(0.,){/Sine}{0}

\FADiagram{(e)}
\FAProp(0.,3.4)(10.,7.)(0.,){/Straight}{-1}
\FAProp(0.,13.)(10.,7.)(0.,){/Straight}{1}
\FAProp(10.,7.)(18.,7.)(0.,){/Sine}{0}
\FALabel(18.,7.6)[t]{$\mathrm x$}
\FAProp(2.5,4.)(12.,3.5)(0.,){/Sine}{0}
\FAProp(4.,4.6)(8.,6.4)(-1.0,){/Sine}{0}

\FADiagram{(f)}
\FAProp(0.,3.4)(10.,7.)(0.,){/Straight}{-1}
\FAProp(0.,13.)(10.,7.)(0.,){/Straight}{1}
\FAProp(10.,7.)(18.,7.)(0.,){/Sine}{0}
\FALabel(18.,7.6)[t]{$\mathrm x$}
\FAProp(4.,4.6)(12.,3.5)(0.,){/Sine}{0}
\FAProp(2.5,4.6)(8.,6.4)(-0.7,){/Sine}{0}

\FADiagram{(g)}
\FAProp(0.,3.4)(10.,7.)(0.,){/Straight}{-1}
\FAProp(0.,13.)(10.,7.)(0.,){/Straight}{1}
\FAProp(10.,7.)(18.,7.)(0.,){/Sine}{0}
\FALabel(18.,7.6)[t]{$\mathrm x$}
\FAProp(2.5,4.)(12.,3.5)(0.,){/Sine}{0}
\FAProp(5.,10.)(5.,5.8)(0.,){/Sine}{0}

\FADiagram{(h)}
\FAProp(0.,3.4)(10.,7.)(0.,){/Straight}{-1}
\FAProp(0.,13.)(10.,7.)(0.,){/Straight}{1}
\FAProp(10.,7.)(18.,7.)(0.,){/Sine}{0}
\FALabel(18.,7.6)[t]{$\mathrm x$}
\FAProp(6.5,5.5)(13.,5.)(0.,){/Sine}{0}
\FAProp(5.,10.)(5.,5.8)(0.,){/Sine}{0}